\documentstyle[12pt,fleqn,cite]{article}
\newcommand{\sect}[1]{\setcounter{equation}{0}\section{#1}}

\def\sxn#1{\sect{#1}}

\def\be{\begin{equation}}
\def\ee{\end{equation}}
\def\bea{\begin{eqnarray}}
\def\eea{\end{eqnarray}}
\def\pa{\partial}

\begin{document}
\begin{titlepage}
\begin{center}
\hfill hep-th/9711040\\
\hfill IMSC/97/09/37\\
\hfill IP/BBSR/97-34\\
\hfill MRI-PHY/p970925\\
\vskip .2in

{\Large \bf Oscillating D-Strings from  IIB Matrix Theory}
\vskip .5in

{\bf Anindya K. Biswas$^{*}$,  Alok Kumar$^{*}$ 
and Gautam Sengupta$^{\dagger}$}\\
\vskip .1in
{\em $*$ Institute of Physics,\\
Bhubaneswar 751 005, INDIA}
\vskip .5cm
{\em $\dagger$ Institute of Mathematical Sciences,\\
C.I.T. Campus, Taramani,\\
Madras 600 113, INDIA\footnote{Address after 
December 15, 1997: Department of Physics, Indian Institute of Technology, 
Kanpur 208 016, INDIA}}
\end{center}

\begin{center} {\bf ABSTRACT}
\end{center}
\begin{quotation}\noindent
\baselineskip 10pt

We present a class of BPS solutions of the IIB 
Matrix Theory which preserve 1/4 supersymmetry. 
The solutions desrcibe D-string configurations 
with left-moving oscillations. We 
demonstrate that the one-loop quantum effective action
of Matrix Theory vanishes for this solution, 
confirming its BPS nature. We also study the 
world-volume gauge theory of oscillating strings 
and show its connection with static D-strings.

\end{quotation}
\vskip .2in
November 1997\\
\end{titlepage}
\vfill
\eject


\sxn{Introduction}

One of the most challenging problem in string theory has 
been to understand its strong coupling 
aspects\cite{senijm,witten95},
including its moduli 
space structure in the full quantum theory at the 
non-perturbative level. One also hopes that
such an investigation will lead to an understanding
of supersymmetry breaking in these theories and will 
give the correct string theory 
vacuum describing the real world. 
As is well known, the unravelling of non-perturbative
aspects includes an analysis of the soliton 
spectrum\cite{schwarz}, and their moduli dependence. 
Investigations along these lines have also 
led to a better understanding of the 
confinement mechanism in supersymmetric gauge theories
from string theory point of view\cite{kachru}. 

A useful mechanism in  
studying the strong coupling aspects has been the 
D-brane constructions of string solitons\cite{polchinski}. 
As a result, one can obtain the soliton spectrum and
their interactions using open string Conformal Field Theory.
Since the D-branes preserve a certain amount of 
supersymmetry, they are stable solitonic superstring 
vacua, around which a quantum field theory of the 
world-volume degrees of freedom can be formulated\cite{banks}. 
Many such D-brane configurations have been 
obtained\cite{gauntlett} 
and the corresponding effective world-volume action have been
analyzed. Among them, of 
particular interest have been some of the six\cite{six} and
two\cite{two} dimensional supersymmetric gauge theories. 

In the Matrix Theory\cite{bfss}\cite{banks,anoth,lowe} 
proposal by Banks,
Fischler, Shenker and Susskind (BFSS), the $SU(N)$ 
$(N\rightarrow \infty)$ world-volume 
gauge theory of $N$ D-branes have 
themselves been conjectured to be a fundamental theory
describing both the perturbative and non-perturbative
aspects of string theory. In simplest cases, these are 
the dimensional reduction of the $D=10$, $N=1$ 
Yang-Mills Theory to the 
relevant world-volume dimension. In this context, 
it has been shown that various 
brane solutions of string theory\cite{anoth,banks}, 
including their charges, 
can be obtained from classical solutions in such gauge 
theories. We will concentrate on the 
IIB Matrix theory\cite{ikkt}
which proposes that the $10$-dimensional type IIB
string theory is described by the dimensional 
reduction of the $D=10$, $N=1$ $SU(N)$ guage theory to 
zero dimension. This 
possesses a  manifest Lorentz invariance. 
The emergence of a D-string from such a 
Matrix theory has also been shown through an 
analysis of their interactions. 
A duality among Matrix theories 
proposed earlier for describing $M$-Theory and the 
one for the type IIB theory has also been argued. 

In this paper, we generalize some of the results in 
\cite{ikkt} and write down an infinite set of 
classical solutions of the 
IIB Matrix theory\cite{makeenko} by 
solving the field equations. These are classical 
gauge field configurations which
correspond to D-strings with chiral (left-moving)
oscillations. The existence of 
these solutions follow from oscillating 
fundamental string solutions\cite{dabol} in type IIB 
string theory and its $SL(2, Z)$ S-duality in 
10-dimensions\cite{schwarz}. 
As in the case of fundamental strings,
we show that the matrix theory solutions preserve 
$1/4$ supersymmetry. 

The BPS mass formula for type IIB string theory, when 
compactified to 9-dimensions, have been written down 
earlier. They are parameterized by integers
$(m, n)$, namely internal momenta and winding 
in the compactified direction, as well as by the gauge 
charges $(p, q)$ corresponding to the NS-NS and 
R-R antisymmetric tensor fields in 10-dimensions. It is 
also known that this BPS formula is invariant under 
the $SL(2, Z)$ $U$-duality in 9-dimensions, which 
follows from the $S$-duality of the 10-dimensional 
type IIB strings. In this paper we 
mainly concentrate on the BPS formula for $(p=0, q=1)$ case 
which corresponds to  a single 
D-string. An explicit form for the BPS formula for
this case can be derived by using the $SL(2, Z)$ 
duality on the mass formula of the fundamental 
IIB string in 9-dimensions  and by restricting 
to the supersymmetric ground states. The mass formula 
for the fundamental string has a form\cite{schwarz}:
\be
  M^2 = ({m\over R_B})^2 + (2 \pi R_B n T_q)^2
		+ 4 \pi T_q (N_L + N_R),
			\label{mass}
\ee
with 
\be
	N_R- N_L = m n,
\ee
and $T_q$ is the string tension of the fundamental string. 
A general formula for the $(p, q)$ string involves a 
generalization in the definition of $T_q$, written in 
terms of the ten-dimensional axion-dilaton moduli as:
\be
	T^2_q = [ p^2 + e^{2\phi_0} (p \chi_0 + q )^2]
			e^{-\phi_0}T^2,
				\label{tension}
\ee  
For a $(0, 1)$ string we then have 
$T_q = e^{{1\over 2}{\phi_0}}T$. 

The BPS states which preserve $1/2$ supersymmetry and 
their interactions have already been analyzed in the 
Matrix Theory context\cite{ikkt}. 
They correspond to the 
supersymmetric ground states $N_L = N_R = 0$. We 
will examine the BPS configurations of the Matrix Theory 
preserving $1/4$ supersymmetry. They are the 
supersymmetric ground states with either $N_L = 0$ or 
$N_R = 0$ and provide a rich spectra parameterized by 
integers $(m, n)$. The BPS mass then satisfies the 
relation:
\be
	M_{BPS} = (2\pi R_B n T_q + m/R_B).
			\label{mbps}
\ee
The mass formula (\ref{mbps}) is an exact 
expression which does not receive quantum corrections. 
In Matrix Theory we verify this by showing that the 
one-loop quantum effective action for our solution 
vanishes. 

As an application of the IIB Matrix Theory, 
we then obtain the world-volume gauge theory
in the classical background of an oscillating D-string 
solution. It is known that the world-volume theory for a static 
D-string configuration of 
IIB matrix theory is a two dimensional 
gauge theory with $(8, 8)$ 
supersymmetry\cite{banks,li}.
In this paper, 
we obtain an explicit expression for the supersymmetric
world-volume gauge theory action 
with $(8, 0)$ supersymmetry from the Matrix Theory action. 
We show its Lorentz, gauge and supersymmetry invariance. 
The gauge and supersymmetry invariance are the 
residual symmetries of the orginal type IIB theory. 
The supersymmetry is a global symmetry in this
case, as it originates from the global supersymmetry of the 
Green-Schwarz superstring action, in the Schild gauge, or
from the supersymmetry of the $N=1$ Yang-Mills theory in 
10-dimensions. The gauge invariance of the $(8, 0)$
world-volume action also follows from that of the gauge 
invariance of the 10-dimensional superYang-Mills theory. 
Although the final model does not possess an explicit
left-right symmetry, we will argue in section-4,
from Matrix Theory point of view, that the 
particle spectrum is anomaly free. We also argue 
that the worldsheet actions for the static and 
oscillating strings define equivalent quantum 
field theories. This is demonstrated through 
a mapping of operators in the
two cases. Physically this also implies that the static string
is a quantum state of the world-volume theory in the
classical background we have studied.  

This work has been partly
motivated by an analysis of BPS states  in compactified 
M-Theory using BFSS model\cite{gopa}. 
We have carried out this analysis
in the framework of $S^1$ compactified IIB 
Matrix Theory\cite{ikkt}. 
The rest of the paper is organized as follows. In 
section-2, we review the oscillating 
fundamental string solutions from supergravity point
of view and mention how the 
corresponding D-strings can be obtained using the 
S-duality of the ten-dimensional type IIB string theory.  
In section-3, we obtain these solutions from  
IIB Matrix theory. We also show that the Matrix Theory 
solution preserves $1/4$ supersymmetry.
In this section, we also point out that the one-loop
quantum effective action of the Matrix Theory, for
this solution, vanishes. In section-4, we present the 
$(8, 0)$ supersymmetric gauge theory 
of oscillating strings and show its 
connection with static strings.  
Conclusions and discussions are presented in section-5.

\sxn{Oscillating String Solution}

We now start with a review of oscillating string 
solution in string theory\cite{dabol}. They were
obtained as a generalization of the static fundamental 
strings found earlier\cite{harvey} and are the solutions of 
the supergravity equations of motion. The 
singularity of the field configuration represents the 
position of the string. 
However unlike the static case, they correspond to 
the states preserving only $1/4$ supersymmetry. 

It is also known that the static fundamental string 
solutions can be identified with charged extremal 
black holes in one lower dimension. Similarly, the
oscillating string solutions, after compactification 
along its length, can asymptotically be identified with the 
supersymmetric, stationary, rotating, charged black holes.
In the context of our discussion in the last section,
the static 
string is a supersymmetric ground state and
the oscillator numbers are fixed to their minimum 
values $N_L = N_R = 0$. On the 
other hand, in the oscillating string configuration 
only $N_R = 0$ and $N_L$ is an arbitrary oscillator number. 
The oscillating string solutions require,
from the space-time point of view,  
the presence of a (large)
compactified direction on which the string is wrapped,
as otherwise the only BPS configurations are those 
preserving $1/2$ supersymmetry in ten noncompact
dimensions. We take $x^1$ as the compactified coordinate
of radius $R$. 

The supergravity solution corresponding to the oscillating
fundamental string is given as:
\bea
	&ds^2  = - e^{2\phi}du dv + [e^{2\phi} p(v) r^{-D+4}
	- (e^{2\phi} - 1) \dot{F}(v)^2 ] dv^2 \cr 
	& + 2 (e^{2\phi} - 1) \dot{\bf F} (v) . d{\bf x} d v
	+ d{\bf x}. d{\bf x},  \cr
        & B_{u v}  = {1\over 2} (e^{2\phi} - 1),  \cr
        & B_{v i} = \dot{F}_i(v)(e^{2\phi} - 1), \cr
	& e^{-2\phi} = 1 + {Q\over {|x-F|^{D-4}}}.  {}
                              \label{solution}
\eea
where, for a fundamental string solution, 
$B_{\mu \nu}$ is  the NS-NS antisymmetric tensor 
field and $F_i (v)$ are
functions of the light-cone coordinate $v = x^0 + x^1$
only. $u = x^0 - x^1$ is the other light-cone coordinate. Dots 
denote the derivative with respect to argument $v$ and 
bold-faced letters denote a vector in the transverse 
directions labeled by indices $i$'s. To match properly with a 
string source, one also requires $p(v)=0 $. The field 
configuration in eqns. (\ref{solution}) define an 
asymptotically flat space. As a result, one can properly
define the ADM mass and charge for the supergravity 
background. It has also been pointed out that  the 
supergravity solution as well as the ADM energy properly 
matches with a string source, written in terms of the
worldsheet coordinates $\tau$ and $\sigma$ as:
\bea
	& V(\tau, \sigma) = 2 R n \sigma^+, &\cr
	& U(\tau, \sigma)  = (2 R n + a) 
	\sigma^- + \int^{V} \dot{F}^2,  \cr
	& {\bf X}(\tau, \sigma)  = {\bf F} (V)  {}, 
		\label{source}
\eea
where $\sigma^{\pm} = \tau \pm \sigma$ and $V$, $U$ are the 
space-time light-cone string coordinates:
$U = X^0 - X^1$, $V= X^0 + X^1$ and  $X^i$ 
are once again the string coordinates
along the transverse directions. The constant `$a$' is the 
zero mode of $\dot{\bf F}^2$:
\be
	a = {1\over \pi} \int_0^{2\pi R n}\dot{\bf F}^2
			\label{defa}, 
\ee
and $F_i$'s have no zero modes. 
The oscillating string is specified by the left-moving wave
profile $F_i (v)$. In \cite{dabol} some specific wave 
profiles have been used to show the connection of the 
oscillating string solution with the charged rotating 
black holes. For our purposes, however, we do not need 
their specific form. 

The worldsheet configuration (\ref{source}) has been 
identified with a string source of momenta and 
winding
\be
p^{\mu} = (2\alpha')^{-1} (2 Rn + a, -a, {\bf 0}),\>\>\>  
n^{\mu} = (0, n, {\bf 0}),
			\label{momenta}
\ee
along the directions
$(X^0, X^1, X^i)$. The internal momenta $m/R$ in the 
compact direction is then specified by integers
\be
	m = - {{R a }\over {2 \alpha'}},
			\label{defm}
\ee
and the oscillator number, obtained by the level-matching 
condition is 
\be
	N_L = {n R a\over 2\alpha'}.
\ee

An oscillating D-string in the supergravity context can be 
obtained by applying an $SL(2, Z)$ duality transformation 
on the fundamental string solution presented above. The 
general procedure, as well as the specific $SL(2, Z)$ 
transformation matrix ($\lambda$) is 
similar to the generation of a static $(p=0, q=1)$ string 
solution from the $(1, 0)$ solution as described in 
\cite{schwarz}. We do not elaborate on them further,
except to note that the fundamental string tension will 
be replaced appropriately by the one for a D-string. 

The string source (\ref{source}) will play a crucial 
role in obtaining a Matrix Theory solution as they, 
with appropriate modifications of string tension will 
specify the gauge field configuration, which are the 
solution of the Matrix Theory field equations. 
So far we have only discussed a single fundamental 
$(1, 0)$ and D-string $(0, 1)$ solutions. The existence
of multiple supersymmetric parallel string configuration 
has also been shown in \cite{dabol}. They correspond to 
$(p, 0)$ and $(0, q)$ type BPS states preserving once
again $1/4$ supersymmetry. 
It may also be possible to obtain higher 
dimensional oscillating branes\cite{russo} 
and to obtain their parallel and orthogonal supersymmetric 
configuration. 

In next section we obtain the oscillating string as a 
solution to the field equation in Matrix Theory and examine its
properties. We also show the BPS nature of
$(0, q)$ or multi D-string solutions of Matrix Theroy 
from the results of one-loop effective action.

\sxn{IIB Matrix Theory}

We now obtain an infinite set of solutions of the 
IIB Matrix theory and show that they correspond to the 
oscillating D-strings discussed in the last 
section from the  supergravity point of view. The IIB
Matrix Theory action is obtained 
by the dimensional reduction of the 
D=10, N=1 $SU(N)$ super Yang-Mills to zero dimension and 
is written as\cite{ikkt}:
\be
	S = \alpha\left( -{1\over 4} Tr [A_{\mu}, A_{\nu}]^2
	- {1\over 2} Tr(\bar{\psi}
	\Gamma^{\mu}[A_{\mu}, \psi])\right)
	+ \beta Tr\>1,
				\label{action}
\ee
where the last term in the action is a ``chemical potential''. 
A similar term in the Schild-type string action is necessary to
show its equivalence with Nambu-Goto action. $\alpha$ and 
$\beta$ are are constants with $\sqrt{\alpha \beta}$ 
defining the D-string tension. Eqn. (\ref{action}) without 
the chemical potential term is also referred as the 
D-instanton Matrix action\cite{makeenko}. 
The constants $\alpha$ and $\beta$ can be determined 
by comparing the string interaction in Matrix Theory with 
those from open strings. The final results are:
$\alpha = {{8\pi^{5\over 2}}\over {\sqrt{3}\gamma}}
{1\over {\alpha'^2 g_s}}$ 
$\beta = {{24\pi^{9\over 2}}\over {\sqrt{3}\gamma}}
{1\over g_s}$, with $\gamma$ being a numerical constant. 

In \cite{ikkt}, the target space metric, represented by the 
the indices $\mu$, has been chosen as Euclidean, whereas the 
oscillating string solutions of \cite{dabol} presented in 
the last section are in the Minkowski metric. We take care
of this discrepancy by putting appropriate factors of `$i$' 
in the solutions of section-2 while computing the one-loop effective 
action. For the moment, however we continue to work with the
Minkowski metric.  

The field equations of the Matrix theory are:
\bea
	[A^{\mu}, [A_{\mu}, A_{\nu}]] & = 0, \cr
	[A_{\mu}, (\Gamma^{\mu}\psi)_{\alpha} ] & = 0. 
			\label{eom}
\eea
As fermions do not have a classical background, 
only the first equation of (\ref{eom}) is cosidered
for analyzing the classical solutions.

The action (\ref{action}) is 
invariant under supersymmetry transformations
\bea
    \delta^{(1)}\psi & = {i\over 2}[A_{\mu}, A_{\nu}]
		\Gamma^{\mu \nu} \epsilon, \>\>\>
    \delta^{(1)}A_{\mu} &= i \bar{\epsilon}
		\Gamma_{\mu} \psi,
				\label{susy1}
\eea
and 
\be
	\delta^{(2)}\psi = \xi, \>\>\>
	\delta^{(2)}A_{\mu} = 0.
			\label{susy2}
\ee
These are also referred as the ``dynamical'' and 
``kinematic'' supersymmetry transformations\cite{banks} and 
follow from the dimensional reduction of the worldsheet 
Green-Schwarz superstring action in Schild
gauge to zero dimension. 
In addition, action (\ref{action}) is invariant under a gauge 
transformation:
\be
	\delta_{gauge} A_{\mu} = i [A_{\mu}, \alpha],\>\>\>
	\delta_{gauge} \psi = i [\psi, \alpha].
			\label{gauge}
\ee

The field equations (\ref{eom}) are now solved by 
infinite dimensional hermitian 
matrices $A_{\mu}$'s. In turn, 
using the familiarity with the quantum mechanics, these 
matrices are represented by the canonicaly conjugate variables,
$q_i$'s and $p_i$'s. The relationship of these solutions
with those in string theory are established through an 
identification of the commutators with the Poisson 
bracket for the Schild action\cite{ikkt}:
\be
	\{X, Y\} = {1\over {\sqrt g}}
	\epsilon^{a b}\pa_a X \pa_b Y,
\ee
where $a$, $b$ denote the worldsheet coordinates
$\tau$, $\sigma$. Moreover one also identifies
\bea
	-i[\>, \>] \rightarrow \{\>, \>\}, \>\>\>
	Tr \rightarrow \int d^2 \sigma \sqrt{g} & ,
	\tau \rightarrow {q\over \sqrt{2 \pi N}}, \>\>\>
	\sigma \rightarrow {p\over \sqrt{2\pi N}}. & {}
		\label{redef}
\eea
with the commutator $[q, p] = 2\pi i$. The static 
D-string:  
\be
	X^{0} = T \tau, \>\>\>
	X^{1} = {L\over 2\pi} \sigma, \>\>\>
	X^{i} = 0,
			\label{static}
\ee
can then be represented by the gauge field configuration:
\be
	A^0 = {T\over \sqrt{2\pi N}} q, \>\>\>
	A^1 = {L\over \sqrt{2\pi N}} p, \>\>\>
	A^i = 0.
			\label{mstatic}
\ee
and satisfies the fields equations (\ref{eom}). 
Similarly the oscillating string can be 
represented by a gauge field configuration which is
obtained through the identifications in (\ref{redef}). 
Continuing to work in light-cone coordinates, the
components $A^{\mu}$'s are given as:
\bea
	& A^V  = 2 Rn \hat{\sigma}^+,  \cr
	& A^U  = (2 R n + a )\hat{\sigma}^- + 
	\int^{\hat{V}} \dot{F}^2, \cr
	& A^i = F^i (\hat{V}), 
			\label{oscil}
\eea 
where $\hat{\sigma}^{\pm} = {{q \pm p}/ \sqrt{2 \pi N}}$
and $\hat{V}$ denotes an operator replacement in the 
function $V$: $\hat{V}(\tau, \sigma)\rightarrow 
V ({q / \sqrt{2\pi N}}, {p/ \sqrt{2\pi N}})$. Once again,
the gauge field configuration for a static string  
(\ref{static}) corresponds to $F^i = 0$ and  $T = L/2\pi = 2 Rn$. 

Now, to verify that $A_{\mu}$'s in eqns. (\ref{oscil})
are solutions of (\ref{eom}), we evaluate their 
commutators. The nonzero ones are:
\bea
      [A^V, A^U] &= -{2i\over N} (2 Rn) (2Rn + a), \cr
	[A^U, A^i] &= {2i\over N} (2Rn) (2 Rn + a) \dot{F}^i. 
		\label{commutator}
\eea
These imply that the field equations are once again 
satisfied. We have therefore found a class of solutions of  the 
Matrix Theory field equations specified by the 
wave-profile ${\bf F} (V)$.

We  now examine the BPS and supersymmetry property of 
the solution (\ref{oscil}). In the background of static
string configuration, the dynamical supersymmetry 
transformation is given as:
\be
	\delta^{(1)}\psi = 
	- {T L\over 2 \pi N} \Gamma^{0 1}\epsilon, 
	\>\>\> \delta^{(1)} A_{\mu} = 0. 
		\label{dyna}
\ee
As a result, only way to preserve some amount of supersymmetry 
is to cancel the dynamical supersymmetry  transformation 
with the kinematic one by 
defining $\xi = \pm {T L\over {2\pi N}}\Gamma^{0 1}\epsilon$. 
We then have $(\delta^1 \pm \delta^2)\psi = 0$ and 
$(\delta^1 \pm \delta^2)A_{\mu} = 0$, which implies that
the solution preserves $1/2$ supersymmetry. 

Now, for the oscillating string background, the dynamical 
supersymmetry transformation can be written as:
\bea
	& \delta^{(1)}\psi  = {1\over {2 N}}  
	(2Rn) (2Rn + a) \left[ \Gamma^{U V} \epsilon
	+ \dot{F}^i \>\Gamma^{V i} \epsilon
				\right],  \cr
	& \delta^{(1)}A_{\mu} = 0, 
			\label{bps}
\eea
Since the transformation $\delta^{(2)}$ is still given by 
eqn.(\ref{susy2}), hence to make sure that a certain amount of
supersymmetry, namely $\delta^{(1)} \pm \delta^{(2)}$, 
is preserved, one also has to impose the 
condition, 
\be
	\dot{F}_i \Gamma^{V i}\epsilon = 0.
				\label{14bps}
\ee	
Before solving this equation explicitly, we notice 
that eqn.(\ref{14bps}) is a chirality condition on 
$\epsilon$ in the light-cone directions, namely
$(1+ \Gamma^0 \Gamma^1 )\epsilon = 0$. Since the string
worldsheet is identified with light-cone coordinates,
eqn.(\ref{14bps})) implies a chirality condition in the
world-volume directions. More explicitly, by choosing  
ten-dimensional 
Gamma matrices in the Majorana representation as:
\bea
	\Gamma^0 = i \pmatrix{ 0 & - I_8 \cr
			I_8 & 0 },\>\>\>
	\Gamma^1 = -i \pmatrix{ 0 & I_8 \cr
			I_8 & 0 },\>\>\>
	\Gamma^i =  \pmatrix{ \gamma^i & 0 \cr
			0 & -\gamma^i },
\eea
and by decomposing the ten-dimensional spinor 
$\epsilon$ in terms of the eight-dimensional ones as
$\epsilon = \pmatrix{\epsilon_L\cr \epsilon_R}$, the 
the condition (\ref{14bps}) implies $\epsilon_R = 0$.
To summarize this part of the discussion, we have shown 
that a cancellation between the ``dynamical'' 
and ``kinematic''
supersymmetry transformations can occur in the
Matrix background (\ref{oscil})
provided half the components
of the dynamical supersymmetry transformations are zero. 
This, in turn, implies that our solution preserves only 
$1/4$ supersymmetry, as expected of an  oscillating 
string. 

To further identify the solution of the Matrix Theory 
(\ref{oscil}) with the oscillating string solution we 
evaluate the classical action for this configuration. We
have:
\be
	S_B = {\alpha\over 2} \left( {{(2Rn)(2Rn +a)}
	\over N}\right)^2 N + \beta N
		\label{classical}
\ee
An extremization with repspect to $N$ and the identification
$\sqrt{\alpha \beta} = 2\pi \rho$, with $\rho$ being the 
string tension now gives 
\be
	S_B = 2 \pi \rho (2 Rn) (2Rn +a).
			\label{value}
\ee 
To verify that the the action (\ref{value}) is proportional to 
the area of the worldsheet, we have directly evaluated the 
Polyakov action, acting as the source for the supergravity 
background, for the oscillating string solution
and shown that it again gives the same value as
in (\ref{value}). Since the solutions in \cite{dabol} 
also satisfy the Virasoro condition, the evaluation of
the Nambu-Goto action in this background also gives the
same value. These results once again confirm that the 
Yang-Mills field configuration do indeed represent the 
oscillating strings, and in turn the infinite hierarchy 
of BPS states. In this context, we notice that the 
BPS mass formula (\ref{mbps}) also follows from the 
time component of the target space momentum 
for these strings written in (\ref{momenta}). 
It is also interesting to note that the Matrix solution
represents a string with well defined string tension for 
generic oscillations $F_i(v)$.  
The change in the value of the action with 
respect to the static string is by an amount:
\be
	\Delta S_B =  2\pi \rho (2Rn). a = 
		{2\over \pi} N_L
			\label{nl}
\ee 
where the last equality follows from the relation
$a = -(2\pi^2 \rho)^{-1} {m\over R}$ for a D-string
which is analogous to (\ref{defm}) for a fundamental 
string through a replacement: 
${1\over {2\pi \alpha'}} \rightarrow 2 \pi \rho$. 
Solutions (\ref{oscil}) can then be interpreted as an 
excitation over the static string state by an 
amount $N_L$ from this point of view as well.

We now analyze the one-loop effective action of the 
Matrix Theory for the classcial background 
(\ref{oscil}) and show that the effective action 
vanishes. The effective action in a general 
background $A_{\mu} = p_{\mu}$ has a form\cite{ikkt}:
\be
	Re W = {1\over 2} Tr log 
	(P^2_{\lambda}\delta_{\mu \nu} - 2 i F_{\mu \nu})
	-{1\over 4} Tr log ((P^2_{\lambda} +
	{i\over 2}F_{\mu \nu}\Gamma^{\mu \nu})
	({{1+\Gamma_{11}}\over 2}))
	- Tr log (P^2_{\lambda}).
			\label{rew}
\ee
where $P_{\mu}$ and $F_{\mu \nu}$ are operators acting on the 
space of matrices as:
\be
	P_{\mu} X = [p_{\mu}, X], \>\>\>
	F_{\mu \nu} X = [f_{\mu \nu}, X],
 			\label{operator}
\ee
with $f_{\mu \nu} = i [p_{\mu}, p_{\nu}]$, $p_{\mu}$
being the operator replacement for variables 
$A_{\mu}$. The terms in 
(\ref{rew}) correspond to the contributions from the 
bosons $A_{\mu}$, the fermions $\psi$ and the 
Fadeev-Popov ghosts repectively. 
It has also been noticed in \cite{ikkt} that the 
imaginary part of $W$ vanishes when $P_i$ is 
zero along at least one of the transverse directions
$i$ and implies the absence of anomaly in the 
world-volume action. 
This holds in our case, provided $F_i = 0$ 
for this index $i$. However it is likely that 
$Im W = 0$ in  generic cases as well.  

To evaluate the effective action in our case, we 
rewrite the gauge field commutators in 
eqn.(\ref{commutator}) in 
Euclidean metric and notice that only nonzero 
components of $F_{\mu \nu}$, namely $F_{0 i}$ and 
$F_{1 i}$ satisfy a  relation: 
$F_{0 i} = - i F_{1 i}$. The form
of the matrix $P^2 \delta_{\mu \nu} - 2 iF_{\mu \nu}$:
\be
	P^2 \delta_{\mu \nu} - 2 i F_{\mu \nu} = 
	\pmatrix{P^2 & 0 & -2i F_{02} & .\cr
		0 & P^2 & -2 i F_{12}& .\cr
		2 i F_{02} & 2 i F_{1 2} & P^2 & 0 \cr
		. & . & 0 & .}
			\label{psquare}
\ee
and property of operators $P^2_{\lambda}$, $F_{\mu \nu}$
in our case: $[P^2_{\lambda}, F_{\mu \nu}] = 0$
then implies that $F_{\mu \nu}$'s cancel out in the 
expression of the determinant of the matix. 
To show this in another way, we expand
\bea
Tr log (P^2_{\lambda} \delta_{\mu \nu} - 2i F_{\mu \nu})= 
	Tr log P^2_{\lambda}\delta_{\mu \nu} + 
	Tr (2 i F_{\mu \nu}/P^2_{\lambda}) + & \cr 
	{1\over 2}(2 i)^2 
	Tr (F_{\mu \alpha}F^{\alpha}_{\nu}/{(P^2_{\lambda})}^2)
	+ ... &\cr 
\eea
and use the fact that only non-vanishing components of 
$\delta_{\mu \nu}$, $F_{\mu \nu}$ in the $u, v$ coordinates
are: $\delta^{u v} = 1$ and $F^{u i} = -2 F_{v i}$. 
It can then be shown that all the higher-order terms vanish, as
one can not form invariants out of the above non-vanishing 
components. 
A similar property of certain classical field configurations, 
namely chiral-null models, have been used to show that they are
an exact solution of the first quantized string 
theory\cite{tset}. We
interesingly observe the appearance of this property in the 
context of Matrix Theory. 

The terms in the 
trace of the matrix $(P^2_{\lambda}
+ {i\over 2} F_{\mu \nu} \Gamma^{\mu \nu})$ 
cancels out similarly.
Various other terms in the effective action (\ref{rew})
then cancel out as in the static case and imply that
the one-loop contribution to the effective action 
for the oscillating case vanishes as well. This 
confirms the exactness of the BPS formula 
(\ref{mbps}) argued on the basis of supersymmetric 
grounds earlier. 

One can also examine the status of the multi-string 
solution. The parallel configuration of oscillating
strings from Matrix Theory can be obtained as 
block-diagonal matrices. Then the cancellations in $W$ occur
within each block in an identical fashion and they once
again vanish, showing that they are BPS configurations as
well.

\sxn{World-Volume Action}

In this section we obtain the 
world-volume gauge theory from IIB Matrix Theory 
for the classical configuration corresponding to 
an oscillating string. We also analyse this
world-volume gauge theory action in some detail and 
show its connection
with static strings upon quantization. 

It is known that the  zero modes of a static 
D-string give rise to an $N=8$ $U(1)$ vector 
multiplet in two dimensions. We will now see 
that the zero modes of an oscillating 
string are the $(8, 0)$ $U(1)$ 
vector multiplets together with 8 scalar multiplets 
containing the world-sheet fermions
of opposite chirality. Similarly zero modes of $N$ 
coinciding D-branes\cite{banks} 
are now expected to give 
rise to an $(8, 0)$ $SU(N)$ gauge theory. Two dimensional 
worldsheet action with $(8, 0)$ and $(4, 0)$ 
supersymmetry have been 
written in other contexts\cite{a80,lowe} 
earlier and it may be interesting
to show the exact connection among these actions.

The world-volume action describing the dynamics of these
fields\cite{banks} can also be obtained 
by adding the quantum fluctuations to the classical 
backgrounds and then by expanding the Matrix Theory action. 
The one-loop effective action of the Matrix Theory
(\ref{rew}) is in fact the quantum effective action of these
gauge theories. Thus in the static case we have\cite{li}:
\bea
      A_0 = -\sigma + \alpha' \tilde{A}_0(\tau, \sigma), &
      A_1 = \tau + \alpha' \tilde{A}_1 (\tau, \sigma), \\
      A_i = \alpha'\phi_i (\tau, \sigma), & 
      \psi = \alpha'\psi (\tau,\sigma), 
				\label{expansion}
\eea
where $\tilde{A}_{\alpha}$ ($\alpha = 0, 1$) now are the
gauge fields on the world-volume whereas the 
transverse components ($\phi_i$'s) are the scalar 
fluctuations. $\psi \equiv \pmatrix{\psi_L\cr \psi_R}$
are the worldsheet fermions which also transform as a 
spinor under an internal $SO(8)$ symmetry. 
These are, as expected, 
the degrees of freedom for an $N=8$ vector multiplet in 
two dimensions and are identified as the bosonic and 
fermionic zero modes of a static string. 

The commutators of Matrix variables, including the 
fluctuations (in the static case) have a form\cite{li}:
\be
	[A_0, A_1] = i \alpha' ( 1 + \alpha' F_{0 1}),\>\>
	[A_{\alpha}, A_i] = i \alpha' D_{\alpha} \phi_i \>\>
	[A_{\alpha}, \psi] = i \alpha' D_{\alpha} \psi,
\ee
where we have used the identification (\ref{redef}) to 
replace the commutators with Poisson Bracket and 
$F_{0 1} = \pa_0 \tilde{A}_1 - \pa_1 \tilde{A}_0 + 
\alpha'\{\tilde{A}_0, \tilde{A}_1\}$ and
$D_{\alpha}\phi^i = \pa_{\alpha}\phi^i + 
\alpha' \{\tilde{A}_{\alpha}, \phi^i\}$.
Then, for a single D-string, the action (\ref{action}) reduces 
to a $U(1)$ gauge theory in two dimensions with 
$N=8$ supersymmetry. The 
bosonic part of the gauge theory action for the $U(1)$ case
has the form:
\be
	S_B = {1\over 2\pi \alpha' g_s}\int d^2 \sigma
	\left( 1 + \alpha'^2 F_{0 1}^2 - \alpha'^2
	D_{\alpha}\phi^i D_{\alpha}\phi^i \right). 
			\label{n8action}
\ee
The first (constant) term in (\ref{n8action}) is the 
contribution of the classical background. In 
Born-Infeld action, they correspond to the term 
involving the  induced world-volume metric. The forms of 
$F_{0 1}$ and $D_{\alpha} \phi_i$ also imply the 
existence of higher (than two) derivative terms in the action. 
These have been identified with the higher order terms
in the expansion of the Born-Infeld action\cite{li}.  
The two derivative terms are the standard 
gauge theory action of the bosonic part of an $N=8$
abelian gauge theory.

We now obtain the worldvolume gauge theory action for the
oscillating configuration from the Matrix Theory and show that 
they correspond to an $(8, 0)$ supersymmetric gauge 
theory in two dimensions. The fact that the solution 
preserves $1/4$ supersymmetry has already been 
pointed out. However this leaves us with two possibilities
for the worldvolume supersymmetry. One can either have
a $(4, 4)$ or an $(8, 0)$ supersymmetric gauge theory in 
two dimensions. The later possibility is more natural in 
our case, as the oscillating string solution discussed 
above is left-right asymmetric. We have however already 
shown the 
breaking of $N=8$ or $(8, 8)$ supersymmtric gauge theory
in two dimensions to an $(8, 0)$ theory explicitly
in eqn.(\ref{14bps}). 

Once again, for writing down the action in 
two dimensions, we expand the Matrix theory fields
around the classical background mentioned above in 
(\ref{oscil}). We now have:
\bea
	A^V &= 2 Rn \hat{\sigma}^+ + \alpha' \tilde{A}^V, \cr
	A^U & = (2Rn + a) \hat{\sigma}^- + 
	\int^{V} {\dot{F}}^2 + \alpha' \tilde{A}^U, \cr
	A^i & = F^i (\hat{V}) + \alpha' \phi^i. 
			\label{expansion2}
\eea
The supersymmetry breaking from $N=8$ or $(8, 8)$ gauge 
theory to an $(8, 0)$ gauge theory can now also be seen 
from the background configuration in 
eqn. (\ref{expansion2}). It is known that the 
R-symmetry for an $N=8$ supersymmetric theory is an
$SO(8)_L \times SO(8)_R$ global symmetry group which
transforms the supercharges as: $(8_v, 1) + (1, 8_v)$. 
Then, due to the background 
configuration for the scalars in eqn. (\ref{expansion2}), 
the left-moving part of the world volume scalars 
acquire vacuum expectation value. This breaks the 
$SO(8)_L \times SO(8)_R$ R-symmetry to  
$SO(8)_L$ and the final worldvolume
theory has an $(8, 0)$ supersymmetry only. 

We now derive this worldvolume action 
and show its invariance under gauge and supersymmetry
transformations.
To write down the worldvolume action, we once again 
compute the commutators appearing in the action (\ref{action})
and make the identifications (\ref{redef}). 
The nonzero ones are:
\bea
	[A^U, A^V]  \rightarrow  2 (2Rn)(2Rn+a) + 
	2 \alpha' (2Rn) \pa_-{\tilde{A}^U} + 2 \alpha'
	[ (2Rn + a) \pa_+\tilde{A^V} - &\cr
	(2Rn)\dot{F}^2 \pa_-\tilde{A^V} ]  
        + \alpha'^2 \{\tilde{A}^U, \tilde{A}^V\} 
	\equiv 2 (2Rn)(2Rn+a) + \alpha' F^{UV}, &
		\label{fuv}
\eea
\bea
	[A^U, A^i]  \rightarrow 2 (2Rn) (2Rn +a)\dot{F}^i 
	+ 2\alpha'[ (2Rn + a) \pa_+ \phi^i
	- (2 Rn) \dot{F}^2 \pa_-\phi^i + & \cr
	 (2Rn) \dot{F}^i \pa_-\tilde{A}^U]
	+ \alpha'^2 \{\tilde{A}^U, \phi^i\} 
	\equiv 2 (2Rn)(2Rn+a)\dot{F}^i + \alpha' D_+\phi^i, &
			\label{dplus}
\eea
\be
	[A^V, A^i] \rightarrow \alpha'[-2 (2Rn)\pa_-\phi^i 
	+ 2 (2Rn) \dot{F}^i \pa_-\tilde{A}^V]
	+ \alpha'^2 \{\tilde{A}^V, \phi^i\}
	\equiv \alpha' D_- \phi^i,
			\label{dminus}
\ee
\be
	[A^i, A^j] \rightarrow 2 \alpha' (2Rn) 
	(\dot{F}^j \pa_-\phi^i - \dot{F}^i \pa_-\phi^j)
	+ \alpha'^2 \{\phi^i, \phi^j\}
	\equiv \alpha' \Phi^{i j},
				\label{phij}
\ee
The bosonic part of the world-volume gauge theory 
action is then obtained
by substituting the above commutators into the bosonic part of
the Matrix Theory action (\ref{action}) and by the 
identifications in eqns. (\ref{fuv})-(\ref{phij}). 
For example, in variables $A_U$, $A_V$ and $A_i$,
the first term in (\ref{action}) has a form:
\be
	S_B = -{\alpha\over 4} Tr \left( 
	- {1\over 2} [A^U, A^V]^2
	+ 2 [A^U, A^i][A^V, A_i]
	+ [A^i, A^j]^2 \right).
		\label{bosonic}
\ee
By ignoring the constant and total derivative terms, 
the bosonic world-volume action is then written as:
\be
	- {\alpha\over 4} \int d^2\sigma 
	{\alpha'^3 \over 2\pi}
	\left[-{1\over 2} {F^{U V}}^2 - 
	2 D_+ \phi^i D_- \phi^i + {\Phi^{i j}}^2 - 
	4 (2Rn)(2Rn+a)\dot{F}^i 
	\{\tilde{A}^V, \phi^i\}\right]
				\label{baction}
\ee
Last term in the above action comes from an expression 
$4 (2Rn) (2Rn +a) \dot{F}^i D_-\phi^i$, 
by dropping the total 
derivative terms. 
The gauge transformations, derived from eqn.(\ref{gauge}), 
in two dimensional gauge theory have a form:
\bea 
	\delta_g \tilde{A}^U  = 2 i (2 Rn + a)\pa_+ \epsilon
	- 2 i (2Rn)\dot{F}^2 \pa_-\epsilon
	+ i \alpha' \{ \tilde{A}^U, \epsilon\},  & \cr
	\delta_g \tilde{A}^V  = - 2i (2Rn) \pa_-\epsilon +
	i \alpha' \{\tilde{A}^V, \epsilon\}, & \cr
	\delta_g \phi^i = - 2i (2Rn) \dot{F}^i \pa_-\epsilon
	+ i \alpha'\{\phi^i, \epsilon\}, & 
			\label{twogauge}
\eea
and imply the following transformations for quantities $F^{UV}$, 
$D_+\phi^i$ and $D_-\phi^i$:
\bea
	\delta_g F^{UV}  = i \alpha' \{F^{UV}, \epsilon\}, & \cr
	\delta_g D_+\phi^i  = -4i (2Rn)(2Rn+a)
	\pa_+\dot{F}^i \pa_-\epsilon
	+ i \alpha'\{D_+\phi_i, \epsilon\}, & \cr
	\delta_g D_-\phi^i  = i \alpha'\{D_-\phi^i, \epsilon\}. &
				\label{gaugefield}
\eea
The gauge invarinace of the action $S_B$ then follows from these 
transformation rules. The fermionic part of the gauge theory 
action can also be written as:
\be
	S_F = - {\alpha \over 2} 
	Tr \left(\bar{\psi}\Gamma^U[A_U, \psi]
	+ \bar{\psi} \Gamma^V[A_V, \psi]
	+ \bar{\psi}\Gamma^i[A_i, \psi] \right). 
			\label{fermion1}
\ee
By using commutators, 
\bea
	& [A^U, \psi] = 2 (2Rn + a) \pa_+ \psi 
	- 2 (2 Rn) \dot{F}^2 \pa_-\psi
	+ \alpha'\{\tilde{A}^U, \psi\}, \cr
	& [A^V, \psi] = - 2 (2Rn) \pa_- \psi 
	+ \alpha'\{\tilde{A}^V, \psi\}, \cr
	& [A^i, \psi] = - 2 (2Rn) \dot{F}^i\pa_-\psi
		+ \alpha'\{\phi^i, \psi\}, 
\eea
and  expanding in terms of the left and right-moving
worldsheet fermions, we have an explicit form:
\bea
	S_F & = -{\alpha \over 2}\int d^2 \sigma 
	{\alpha'^2 \over 2\pi} \left[
	 2 (2Rn+a)\psi_R^T \pa_+ \psi_R
	- 2 (2Rn)\dot{F}^2 \psi_R^T \pa_-\psi_R + 
	\alpha' \psi_R^T\{\tilde{A}^U, \psi_R\} 
			\right.  \cr
	 & - 2 (2Rn)\psi_L^T \pa_- \psi_L + 
	\alpha'\psi_L^T\{\tilde{A}^V, \psi_L\}
	+ 2i (2Rn)\dot{F}^i (\psi_R^T\gamma^i\pa_-\psi_L 
	+ \psi_L^T \gamma^i \pa_- \psi_R)  \cr 
 	& - i \alpha' (\psi_R^T \gamma^i \{\phi_i, \psi_L\}
	+ \psi_L^T\gamma^i\{\phi^i, \psi_R\})
\left.        \right]. 
	\label{fermion2}
\eea

To obtain the supersymmetry transformations for the
two dimensional gauge theory action, given by 
$S= S_B + S_F$, from Matrix Theory,  
we use the condition $\epsilon_R = 0$ 
which follows from (\ref{14bps}). For supersymmetry 
transformtion $\delta = \delta^1 -\delta^2$ we have  
\be
	\delta\tilde{A}^U = 2 i\epsilon_L^T\psi_L, \>\>\>
	\delta \tilde{A}^V = 0, \>\>\>
	\delta \phi^i = - \epsilon_L^T
			\gamma^i \psi_R
\ee
and 
\be
	\delta \psi_L  = {i \over 2}
	\left( F^{UV} + \Phi^{ij}\gamma^{ij}
	\right)\epsilon_L, \>\>\>
	 \delta \psi_R  =  D_-\phi_i \gamma^i \epsilon_L. 
\ee
The supersymmetry invariance of the action 
can then be verified explicitly. In a compact 
(covariant) form, the supersymmetry transformations have an 
explicit form:
\be
	\delta S_B = -\delta S_F = 
	-i Tr [A_{\alpha}, A_{\beta}] [A^{\alpha}, 
	\bar{\epsilon}\Gamma^{\beta}\psi]. 
\ee
A more explicit form of these transformations in terms of 
field variables $\tilde{A^U}$, $\tilde{A^V}$, $\phi^i$ and 
$\psi$ can be written down by using 
eqns. (\ref{fuv})-(\ref{phij}) and identifications
(\ref{redef}). Finally, Lorentz invariance of 
the worldvolume action 
can be seen from the scaling transformations:
\be
	\sigma^+ \rightarrow \lambda \sigma^+, \>\>\>
	\sigma^- \rightarrow \lambda^{-1} \sigma^-, \>\>\>
			\label{lorentz}
\ee
together with the transformation of the bosonic fields:
\be
	\tilde{A}^U \rightarrow \lambda^{-1} 
	\tilde{A}^U, \>\>\>
	\tilde{A}^V \rightarrow \lambda 
	\tilde{A}^V, \>\>\>
	\phi^i \rightarrow \phi^i,\>\>\>
	\dot{F}^i \rightarrow \lambda^{-1} \dot{F}^i
\ee
and those of fermions:
\be
	\psi_R \rightarrow \lambda^{1\over 2}\psi_R,\>\>\>
	\psi_L \rightarrow \lambda^{-{1\over 2}}\psi_L. 
\ee

We have already pointed out in section-3 that the world-volume
action is anomaly free. This is essentially due to the fact that 
spectrum contains equal number of left 
and right moving fermions. Moreover, 
as in the case of gauge theory with $(8, 8)$ 
supersymmetry and the corresponding Abelian 
Born-Infeld action, all the fermions
as well as matter scalars are neutral under  
gauge symmetry on the world-volume for the oscillating
D-string as well. As a result they
do not contribute to the anomaly. 
In section-3 we have argued this more concretely
by pointing out that 
$Im W =0$ whenever transverse oscillation 
is absent along one of the directions. 

We now discuss the connection of our solution with 
static D-strings by arguing that one can obtain the 
particle spectrum of static strings, from that of the
classical configuration discussed above, after quantization. 
In two dimensions, this implies the quantum equivalence of 
the worldvolume action (\ref{n8action}) for the static case with 
the oscillating one (\ref{bosonic}) and  (\ref{fermion2}). 
This is true in 
spite of the fact that the manifest symmetries of the 
two actions are quite different. However this is expected 
from a different angle, namely the absence of spontaneous 
symmetry breaking in two dimensions. Our results of this
part therefore imply that the collective modes of 
oscillating string have, in their spectrum, the 
static string as well. 

To analyse the quantum spectrum corresponding to the
action (\ref{bosonic}) and (\ref{fermion2}), we set the 
gauge field fluctuations to zero and make the choice:
$F^i =0$, for $i\neq 1$ and $F^1 = F$. Furthermore, 
we restrict the analysis to the bosonic sector 
only, as the fermionic part can also be analysed in a 
similar manner. 
After these simplifying assumptions, the
bosonic action for fields $\phi^i$ $(i=2,...,8)$ 
are those of a free field. The field 
$\phi\equiv \phi^1$ is described by an action:
\be
	S^1_B = (-{\alpha\over 4})(2Rn)
	\int d^2 \sigma \left[ (2 Rn +a) \pa_+\phi
	\pa_-\phi - (2Rn)\dot{F}^2 (\pa_-\phi)^2\right]
				\label{simpleaction}
\ee
The equation of motion corresponding to the action
(\ref{simpleaction}) can be written as $\pa_- J^- = 0$, 
where
\be
	J^- = (2Rn + a)\pa_+\phi - (2Rn)\dot{F}^2\pa_-\phi.
			\label{jminus}
\ee
$J^-$ is a chiral conserved current in 
the theory. The equation of motion 
for closed strings can be solved as:
\bea
	\phi = p_L\sigma^+ 
	+ p_R (\sigma^- + {1\over {(2Rn +a)}}
	\int^V \dot{F}^2) + \sum_{m\neq 0} 
	{({\alpha_{-m}\over m})}e^{2i m \sigma^+}
	& \cr
	+ \sum_{m\neq 0}
	{({\tilde{\alpha}_{- m}\over m})}
	e^{- 2i {m\over (2Rn)}[(2Rn+a)\sigma^-
	+ \int^V \dot{F}^2 ]}.	& 
		\label{fsoln.}
\eea  
The canonical formulation can be applied to the
time and space-dependent Lagrangian  
(\ref{simpleaction}) in a standard way 
and leads to a Hamiltonian density of the  
form:
\be
	{\cal{H}} = (-{\alpha\over 4}){(2Rn)\over {2 (2Rn +a)}} 
	{\left({J^-}^2
	+ [(2Rn+a)^2 - 
	(2Rn)^2 \dot{F}^4] (\pa_-\phi)^2\right)}.
		\label{density}
\ee
In writing down equation (\ref{density}), we have 
replaced the canonical momentum by the space and 
time derivatives, to present a simple form. 

To show the mapping of the spectrum of oscillating
string, specified by the oscillators $\alpha$ and 
$\tilde{\alpha}$ with that of the static strings, 
specified by $\bar{\alpha}$, 
$\tilde{\bar{\alpha}}$, we choose
a specific wave-profile $F(v)$ of the 
classical oscillating string solution,
namely $F(v) = \sqrt{a (2Rn)}\left( [sin(V/2Rn) + 
cos (V/2Rn)] + {2\over {2 \pi Rn}}\right)$. This choice 
satisfies the condition that $F(v)$ has no zero mode, 
namely $\int^{2\pi Rn} F(v) = 0$. Then, by 
representing the oscillators
$\alpha$ and $\tilde{\alpha}$ as Fourier components of 
two commuting operators $J^-$ and $\pa_-\phi$ 
respectively and comparing them with the Fourier 
components of similar operators in the static case:
$F=0$, it can be shown that, at $\tau = 0$, the oscillators
are mapped as:
\be
	\bar{\alpha}_m = {(2Rn+a)\over (2Rn)}\alpha_m, \>\>\>
	\tilde{\bar{\alpha}}_m = {(2Rn+a)\over (2Rn)}
	\sum_q J_{m-q}({2aq\over Rn})
	\tilde{\alpha}_q,  
			\label{bassel}
\ee
where $J_m(x)$ are the Bessel functions. Equation
(\ref{bassel}) gives a mapping between  
operators in the spectrum of oscillating and static
strings. Using the above relationship between the
operators $J^-$ and $\pa_-\phi$ in the two cases, 
the solutions for fields themselves can be shown to  
be  related to each other.  
The mapping at $\tau \neq 0$ is given 
by the time-evolution of these operators with respect to 
the corresponding Hamiltonians. 
A similar mapping should be
possible for the fermionic oscillators as well.

\sxn{Conclusions}

We have presented a class of solutions of the 
IIB Matrix theory and shown that the solutions 
preserve $1/4$ supersymmetry. The supersymmetry that is
preserved is chiral in nature in terms of the wave motion
on the D-string. We have confirmed the BPS nature of
these  solutions by computing the one-loop
effective action and derived the world-volume gauge theory. 
It was also shown that the world-volume action in the classical
background of oscillating string is anomaly free for a 
large class of models. However, it 
should be possible to show this property without making any 
assumption about the form of the transverse oscillations.

There can be several applications and generalizations of these
results. First, it will be interesting to extend the 
results of this paper to other extended objects with 
oscillations. A membrane solution of this type has 
already been known\cite{russo} and 
implies that a similar analysis in the BFSS Matrix Theory should 
be possible. Another interesting aspect of this analysis may
be to examine the gauge theories that might arise through 
other oscillating D-brane configurations. The BPS states of strings 
have been analyzed using the results in four 
dimensional gauge theories\cite{gopa} in the context of 
BFSS Matrix Theory. However it should be possible to 
carry out our analysis in similar circumstances. 
One may also be able to apply the results of this paper to 
study black holes in the IIB Matrix Theory picture.
This can be done through the identification of the 
compactified oscillating strings with extremal black holes.

\sxn{Acknowledgements}

We thank D. Jatkar, S. Mukhopadhyaya, S. Naik, S. Panda,  
K. Rama and  A. Sen for invaluable discussions,
suggestions and critical comments. 
G.S. thanks members of the String Theory 
Journal Club of the Institute of Mathematical Sciences, Madras 
for general discussions 
regarding Matrix Theory. We also thank Mehta Research Institute, 
Allahabad for its hospitality where this work was done.  

\vfil
\eject

\end{document}